\documentclass[10pt]{emulateapj}
\usepackage{apjfonts}
\usepackage{graphicx}
\usepackage{psfig}
\slugcomment{{\sc Submitted to ApJL:} December 1, 2009}
\newcommand{\begit}{\begin{itemize}}
\newcommand{\enit}{\end{itemize}}
\newcommand{\begen}{\begin{enumerate}}
\newcommand{\enen}{\end{enumerate}}

\newcommand{\beq}{\begin{equation}}
\newcommand{\eeq}{\end{equation}}
\newcommand{\beqa}{\begin{eqnarray}} 
\newcommand{\eeqa}{\end{eqnarray}} 

\begin{document}

\title{A High Rate of White Dwarf $-$ Neutron Star Mergers 
\& Their Transients}

\author{Todd A.~Thompson\altaffilmark{1,2,4},
Matthew D.~Kistler\altaffilmark{2,3}, \& K.~Z.~Stanek\altaffilmark{1,2}}

\altaffiltext{1}{Department of Astronomy,
The Ohio State University, Columbus, Ohio 43210, USA
thompson@astronomy.ohio-state.edu}
\altaffiltext{2}{Center for Cosmology \& Astro-Particle Physics,
The Ohio State University, Columbus, Ohio 43210, USA}
\altaffiltext{3}{Department of Physics,
The Ohio State University, Columbus, Ohio 43210, USA}
\altaffiltext{4}{Alfred P.~Sloan Fellow}

\begin{abstract}

We argue that the recent groundbreaking discovery by 
Badenes et al.~(2009) of a nearby ($D\approx50$\,pc) white dwarf-neutron star 
(or black hole) binary (SDSS 1257+5428) with a merger timescale $\lesssim500$\,Myr 
implies that such systems are common; we estimate that there are of 
order $\sim10^6$ in the Galaxy. Although subject to large uncertainties, 
the nominal derived merger rate is $\Gamma_{\rm MW}\sim5\times10^{-4}$\,yr$^{-1}$ 
in the Milky Way, just $\sim3-6$ and $\sim20-40$ times less than the 
Type Ia and core-collapse supernova (SN) rates, respectively.  
This implies that the merger rate is $\sim0.5-1\times10^4$\,Gpc$^{-3}$ 
yr$^{-1}$ in the local universe, $\sim5000-10000$ times more than the 
observed (beaming-uncorrected) long-duration gamma-ray burst (GRB) rate.  
We estimate the lower limit on the rate in the Galaxy to be 
$\Gamma_{\rm MW}\gtrsim2.5\times10^{-5}$\,yr$^{-1}$ at 95\%
confidence. We briefly discuss the implications of this finding
for the census of long- and short-duration GRBs and their 
progenitors, the frequency of tight binary companions to Type-Ib/c 
SN progenitors, the origin of ultra-high energy cosmic rays (UHECRs),
the formation of rapidly rotating neutron stars and 
$\sim2-3$\,M$_\odot$ black holes, 
the census of faint Ia-like SNe, as well as for 
upcoming and current transient surveys (e.g., LOSS, PTF, LSST),
and for high- (LIGO) and low-frequency (LISA) gravitational wave searches.

\end{abstract}


\section{Introduction: The Argument}
\label{section:introduction}

The recent paper by Badenes et al.~(2009) describes their discovery as
part of the SWARMS survey of  
a probable white dwarf-neutron star/black hole (WD-NS/BH)\footnote{
We adopt this terminology since it is not known if the unseen companion is 
a NS or BH. Badenes et al.~(2009) discuss a third option ---
that the companion is another unseen WD --- but they do not favor this
option. We return to this issue in \S\ref{section:implications}.}
binary (SDSS 1257+5428) at a distance of $D\approx48^{+10}_{-19}$\,pc, 
with orbital period of $\approx4.6$\,hr, and radial velocity semi-amplitude 
of  $\approx320$\,km s$^{-1}$.  The WD is of spectral
type DA, has a $g$-band magnitude of 16.8, and has a cooling age of 
$t_{\rm cool}\approx2.0\pm1.0$\,Gyr.
The system masses are $M_{\rm WD}\simeq0.92^{+0.28}_{-0.32}$ and 
$M_{\rm NS/BH}\sin(i)\simeq1.62^{+0.20}_{-0.25}$\,M$_\odot$. 
The transverse velocity of the system is $\approx11$\,km s$^{-1}$
and the total spatial velocity is plausibly $\sim120$\,km s$^{-1}$.
For $i=60^{\circ}$, the semi-major axis is $\approx0.01$\,AU.
The orbital period and masses imply a merger timescale for the system of 
$t_{\rm merge} \leq 511^{+342}_{-141}$\,Myr and for an assumed inclination
angle of $i=60^{\circ}$, $t_{\rm merge} \sim 267^{+165}_{-70}$\,Myr.

Taken at face value, the detection of this remarkable binary 
implies that the number of such  systems in the Galaxy is very 
large.  The WD has an apparent $g$-band magnitude $\approx2$
magnitudes brighter than the limiting magnitude of the survey,
18.9 (Badenes et al.~2009; Mullally et al.~2009).  Thus, the 
volume probed by the survey for systems analogous to SDSS 1257+5428 is
$V_{\rm SWARMS}\sim f_{\rm SDSS}(4\pi/3)(125\,{\rm pc})^3\sim2\times10^6$\,pc$^3$,
where $f_{\rm SDSS}\sim1/4$ is the fraction of the sky probed by SDSS to $g\approx18.9$
in SWARMS, and where we have taken $f_{\rm SDSS}(4\pi/3)\sim1$. 
The total volume of the Galaxy is $V_{\rm MW}\sim2\pi r^2 h$, where 
$r\sim8.5r_{\rm 8,5\,kpc}$\,kpc is the galacto-centric radius and 
$h\sim 4 h_{\rm 4\,kpc}$\,kpc
is the scale height accessible to objects with total space motion of 
$\sim100$\,km s$^{-1}$.  Thus,
\beq
\frac{V_{\rm SWARMS}}{V_{\rm MW}}
\sim1\times10^{-6}\,r_{8.5\,\rm kpc}^{-2} h_{4\,\rm kpc}^{-1}.
\label{Vratio}
\eeq
The primary point of this {\it Letter} is that this ratio 
is exceedingly small.  Since equation (\ref{Vratio}) is equivalent to the 
probability of detecting such a system if it was the only such 
system in the Galaxy, we conclude that it is not the only such system.  
More probable is that each of the 
$V_{\rm SWARMS}$-sized volumes that 
constitute the Galaxy is populated by $N\gtrsim1$ such 
binaries.  The total number of WD-NS/BH binaries in the 
Galaxy is then\footnote{We note that 
the density of stars in the Solar neighborhood can be estimated 
similarly, and with at least order-of-magnitude fidelity, from the 
first measured parallax of 61 Cygni by Bessel (1838).}
\beqa
N_{\rm MW}\sim N\left(\frac{V_{\rm MW}}{V_{\rm SWARMS}}\right)
\sim10^6 N\,\,r_{8.5\,\rm kpc}^{2} h_{4\,\rm kpc}
\label{number}
\eeqa
and the current merger rate is 
\beqa
\Gamma_{\rm MW}
\sim\frac{N_{\rm MW}}{t_{\rm life}}
\sim 5\times10^{-4}N\,{\rm yr^{-1}}\,\,r_{8.5\,\rm kpc}^{2} 
h_{4\,\rm kpc}\,t_{\rm 2Gyr}^{-1},
\label{rate}
\eeqa
where $t_{\rm 2\,Gyr}=t_{\rm life}/2$\,Gyr and $t_{\rm life}=t_{\rm cool}+t_{\rm merge}$
(e.g., Phinney 1991).

We caution that the estimate of the number of such systems in the Galaxy in equation (\ref{number})
and the merger rate in equation (\ref{rate})
is highly uncertain.  Ignoring the uncertainties in the ratio of volumes and
in $t_{\rm life}$, the Poisson error on $N$, given the single 
detection, yields a lower limit of $N\gtrsim0.05$ 
($N_{\rm MW}\gtrsim5\times10^4$; $\Gamma_{\rm MW}\gtrsim2.5\times10^{-5}$\,yr$^{-1}$) and 
$\gtrsim0.01$ ($N_{\rm MW}\gtrsim1\times10^4$;$\Gamma_{\rm MW}\gtrsim5\times10^{-6}$\,yr$^{-1}$) 
at 95\% and 99\% confidence, respectively (e.g., Gehrels 1986).
A single additional such binary detected in the local 
$\sim10^6$\,pc$^3$ would provide strong evidence for $N\sim1$.  
Unfortunately,  the lack of such a future detection in SWARMS 
does not strongly constrain $\Gamma_{\rm MW}$.\footnote{Based on 
the constraints on the mass function alone, there is a 
significant chance ($\sim50$\.\%) that one of the systems recently 
reported by Kilic et al.~(2009) is also a WD-NS binary.  
Since the survey volume is much larger for the Kilic et al.~survey
(they target bright Helium WDs), if the system is confirmed 
to be a WD-NS binary, $\Gamma_{\rm MW}$ is still likely to 
be dominated by SDSS 1257+5428.}
In addition to the 
uncertainty in $N$, there is uncertainty at the factor of $\sim2-4$ 
level in $V_{\rm MW}$; however, our fiducial values for $r$ and $h$ 
are reasonable, given the age and space velocity of the binary 
considered.  Note also that SWARMS selects for edge-on
binaries due to the low resolution of the SDSS spectra
used to target interesting WDs via radial velocity variations
(Badenes et al.~2009; Mullally et al.~2009).  This effect
may increase the inferred rate by a factor of order $\sim2-4$.
The ambiguity in the cooling age of SDSS 1257+5428 adds 
an additional factor of $\sim2-4$ uncertainty.  Finally,
our estimate tacitly assumes that the binary is equally 
detectable by SWARMS over its few Gyr lifetime from 
birth to death; the fact that the WD was brighter in the 
past implies a lower overall rate than given in equation (\ref{rate}), 
but the magnitude of this correction depends on the luminosity function of WDs 
in such binaries.

\section{Discussion: The Rate}
\label{section:discussion}

The nominal estimate of the merger rate for WD-NS/BH binaries
in equation (\ref{rate}) is very high.  Taking the 
core-collapse supernova (SN) rate to be $\sim1-2\times10^{-2}$\,yr$^{-1}$ 
in the Galaxy, we see that the WD-NS/BH merger rate is $\sim20-40$ times 
smaller.  Comparison with the observed rate of core-collapse SNe in the local
100\,Mpc volume (see, e.g., the compilation of data by Horiuchi et al.~2009),
$\sim1-2\times10^5$\,Gpc$^{-3}$ yr$^{-1}$, implies that 
the average rate of WD-NS/BH mergers in the local universe is 
$\sim0.5-1\times10^4$\,Gpc$^{-3}$ yr$^{-1}$.  This is a 
factor of just $\sim3-6$ lower than the local Type Ia rate, 
$\sim3\times10^4$\,Gpc$^{-3}$ yr$^{-1}$ (Dilday et al.~2008).
See Table \ref{table}.

Although the rates of short- and long-duration 
gamma-ray bursts (GRBs) are uncertain because of the overall beaming correction
for the two populations, $\Gamma_{\rm MW}$ in equation (\ref{rate}) 
is at least $\sim10^2$ times larger than the rates inferred for either population;
the long-duration GRB rate is observed to be $\sim1$\,Gpc$^{-3}$ yr$^{-1}$,
and the beaming-corrected rate is $\sim30$\,Gpc$^{-3}$ yr$^{-1}$ (e.g., Guetta et al.~2005).
The rate of short-duration GRBs is also small ($\sim10$\,Gpc$^{-3}$ yr$^{-1}$; Nakar 2007).  
This disparity between $\Gamma_{\rm MW}$ and the GRB rate is particularly
interesting because these mergers have been suggested as a mechanism
for some long- and short-duration GRBs (\S\ref{section:implications}).
Interestingly, our rate for $\Gamma_{\rm MW}$ is only $\sim5$ times 
higher than the current estimate for the rate of NS-NS mergers in the 
Galaxy (Kalogera et al.~2004ab), often thought to be associated
with short-duration GRBs (see Table \ref{table}).

Our estimate of $\Gamma_{\rm MW}$ is also significantly larger than  
the rate derived from statistical methods based on the 
3 previously known WD-NS(pulsar) binaries that will merge
within a Hubble time,
$\sim0.2-10\times10^{-6}$\,yr$^{-1}$,  obtained without making 
an overall correction for pulsar beaming (Kim et al.~2004).  
This result is statistically dominated by the single WD-NS 
binary system PSR J1141-6545, whose merger timescale is similar 
to the system described here. The rate derived in 
Kim et al.~(2004) is about an order of magnitude smaller 
than that obtained from population synthesis studies (e.g., 
Kalogera et al.~2005), and it is about $\sim10^2$ times 
smaller than $\Gamma_{\rm MW}$ in equation (\ref{rate}).  
The estimate by Fryer et al.~(1999a) is similarly small,
but with large uncertainties.

Our estimate of $\Gamma_{\rm MW}$ is $\sim1-10$ times higher
than the rate advocated by Davies et al.~(2002) in their 
binary evolution scenario developed to explain the WD-pulsar 
systems J1141-6545 and B2303+46.\footnote{Note that their 
production rate of tight binaries is highest (closest to 
eq.~\ref{rate}) for a flat (i.e., not Salpeter) distribution 
of secondary masses and for large values of the common envelope 
evolution parameter.}  In their model, the initial primary 
transfers enough of its mass to the secondary that the latter 
becomes the more massive star.  The core of the original primary 
makes the WD, which enters a common envelope with the more 
massive secondary when it evolves off the main sequence.  In the
case of their PSR J1141-6545-like binaries,
the secondary helium star evolves to contact
with the primary and the majority of its envelope is ejected from 
the system.  The secondary then undergoes core collapse, producing
a NS.  However, since SDSS 1257+5428 is not highly 
eccentric it is unclear to what extent it may be viewed as 
analogous to J1141-6545 (see [10], \S\ref{section:implications}).

Our 95\% confidence lower limit on the rate --- 
$\Gamma_{\rm MW}\gtrsim2.5\times10^{-5}$\,yr$^{-1}$ --- is still 
larger than either of the observed or beaming-corrected short- or 
long-duration GRB rates, but overlaps with the probable distribution 
of NS-NS merger rates for the Galaxy (Kalogera et al.~2004ab), the 
upper end of the pulsar beaming-corrected rates derived from PSR 
J1141-6545 (Kim et al.~2004), the rates inferred from population 
synthesis studies (Davies et al.~2002; Kalogera et al.~2005),
the rate derived by Nelemans et al.~(2001), and the estimate 
of Edwards \& Bailes (2001) of $\sim10^{-5}$\,yr$^{-1}$ based on 
PSR J1141-6545.

%%%%%%%%%%%%%%%%%%%%%%%%%%%%%%%%%%%%%%%%%%%%%%%%%%%%%%%%%%%%%%%%
\begin{table}[!t]
\begin{center}
\caption{Events \& Rates
\label{table}}
\begin{tabular}{lccccccccc}
\hline \hline
\\
\multicolumn{1}{c}{Event Type} &
\multicolumn{1}{c}{Rate} &
\multicolumn{1}{c}{Reference} \\

\multicolumn{1}{c}{} &
\multicolumn{1}{c}{ (10$^4$\,Gpc$^{-3}$ yr$^{-1}$)} &
\multicolumn{1}{c}{} \\
\\
\hline
\hline
\\
WD-NS/BH Merger & $0.5-1$      &   this work \\
\\
\hline
\\
Type II SN   & $10-20$     &   Horiuchi et al.~(2009) \\
Type Ia SN   & $3$         &   Dilday et al.~(2008) \\
NS-NS Merger & $0.1$\tablenotemark{a}       &   Kalogera et al.~(2004ab) \\
Short GRB    & \,\,$0.001$\tablenotemark{b,c}     &   Nakar (2007) \\
Long GRB     & $0.0001$\tablenotemark{b}     &   Guetta et al.~(2005) \\
\\
\hline
\hline
\end{tabular}
\tablenotetext{a}{Calculated using the rate $\approx10^{-4}$\,yr$^{-1}$ derived in 
Kalogera et al.~(2004ab) and the conversion factor of $10^{-2}$\,Mpc$^{-3}$ based on the
$B$-band luminosity of the local universe from 
Kalogera et al.~(2001), their Section 4.}
\tablenotetext{b}{Uncorrected for beaming.}
\tablenotetext{c}{Short-duration GRBs with extended emission (e.g., 060614)
amount to $\sim25$\% of the total short-duration GRB population (Norris et al.~2009).}
\end{center}
\end{table}
%%%%%%%%%%%%%%%%%%%%%%%%%%%%%%%%%%%%%%%%%%%%%%%%%%%%%%%%%%%%%%%%

\section{Implications}
\label{section:implications}

The estimate of the merger rate in equation (\ref{rate})
has a number of implications:

\noindent {\it 1.~GRBs \& Transients:}
The merger of WD-NS/BH binaries has been discussed as a mechanism 
for producing long-duration GRBs (e.g., Fryer et al.~1999b)
and the new class of short-duration GRBs with extended emission (e.g., 
GRB 060614; King et al.~2007; see Gehrels et al.~2006 \& Gal-Yam et al.~2006).
A recent study by Norris et al.~(2009) finds that only $\sim25$\% of short 
duration GRBs have extended emission.  Thus, if King et al.~(2007) are
correct and WD-NS mergers are the central engine for 060614-like bursts,
then either equation (\ref{rate})  overestimates the rate by a factor
of $\sim10^2-10^3$, only a very small percentage of WD-NS/BH mergers produce
GRBs, or these events are very highly 
beamed --- with jet opening angles many times smaller than most estimates
(e.g., Nakar 2007; Guetta et al.~2005).  
Similar statements can be made for short-duration GRBs without extended emission,
or for long-duration GRBs unaccompanied by SNe (see Table \ref{table}).  The natural 
timescale for such transients is long ($\gtrsim100$\,s) since the WD will be tidally 
disrupted at a radial distance from the NS comparable to its
own radius.   Equation (\ref{rate}) implies that transients 
generated by WD-NS/BH mergers should be relatively common in 
Milky Way-like  galaxies ($\sim10-30$\% of the Ia rate), 
contrary to what is observed for long-duration GRBs 
(Stanek et al.~2006). 
Whatever transients these mergers produce (see [4] below),
they should be associated with relatively old ($\sim$\,Gyr)
stellar populations, may occur many kpc from their 
host galaxies, and should not correlate with recent star formation.

\noindent {\it 2.~Potential Neutrino Signature:} 
Whether the unseen companion in SDSS 1257+5428 is a NS 
or BH, the merger may generate an interesting neutrino
signature, with a maximum possible total energy radiated of
$\sim3\times10^{53}$\,ergs (similar to 
a NS-producing core-collapse SN).  However,
because the accretion rate is likely to be modest, 
$\lesssim10^{-2}$\,M$_\odot$ s$^{-1}$,  the merger disk is 
unlikely to be radiatively efficient in neutrinos 
(e.g., Popham et al.~1998; Chen \& Beloborodov 2007).
Since BH formation is a likely outcome of the merger
(see [6] below), a distinct signature of this NS-to-BH 
transformation may be seen in the neutrino (or potentially high-energy photon)
lightcurve.

\noindent {\it 3.~Nucleosynthesis:}
The outflow produced from the 
NS/BH during the accretion phase of the WD --- either in a
jet or a wind --- may produce thermodynamic conditions favorable for 
interesting nucleosynthesis.  For very high accretion 
rates ($\sim1$\,M$_\odot$ s$^{-1}$; Chen \& Beloborodov 2007; Metzger et al.~2009)
the outflow may become neutron rich, producing conditions
favorable for production of the $r$-process nuclei (e.g., Metzger et al.~2007).
Even in the absence of a neutron excess (as is more likely 
for the low accretion rates one estimates for
such a merger/disk) if the dynamical timescale for ejection is 
short enough, the $r$-process might still occur (Meyer 2002).
Since the rate estimated in equation (\ref{rate}) 
is relatively high with respect to NS-NS mergers
(a commonly discussed $r$-process production site; Freiburghaus et al.~1999), 
WD-NS/BH mergers may make an interesting contribution
to the heavy-nucleus budget of the Galaxy.

\noindent {\it 4.~Optical Signature:}
Similar to the case of WD-WD mergers, or accretion-induced
collapse of a WD to a NS as a result of accretion from a companion,
as considered in Metzger et al.~(2009), there may be a low-luminosity 
optical transient generated by $\lesssim10^{-2}$\,M$_\odot$ of $^{56}$Ni
associated with the merger of WD-NS binaries.  This material may
be ejected as an initially proton-rich outflow from the disk around the
NS or BH. The total ejecta
mass would be small on the scale of normal SNe. 
The rate in equation (\ref{rate}) is small enough that 
such transients may be so far unknown, but we speculate that they 
might be linked to classes of unusual SNe like SN 1991bg 
(Filippenko et al.~1992),
2002bj (Poznanski et al.~2009), 2008ha (Foley et al.~2009; Valenti et al.~2009), 
or 2005cz (Kawabata et al.~2009) and 2005E (Perets et al.~2009). 
Such events should be relatively
common in the local universe ($\sim10-30$\% of the Ia rate), 
trace relatively old stellar populations, have no optically-visible 
progenitors, be seen many kpc from their host galaxies (e.g., SN 2005E), and 
many should be seen by current transient survey efforts such as
LOSS (Li et al.~2000),
Palomar Transient Factory (PTF) and the Panoramic Survey Telescope and Rapid Response 
System (PanSTARRS); the future Large Synoptic Survey Telescope 
(LSST) should see many hundreds per year.  The composition of the 
ejecta may vary, depending on the type of WD disrupted (e.g., He, C/O, O/Ne/Mg).

\noindent {\it 5.~Gravitational Waves:}
Assuming $M_{\rm WD}=0.9$\,M$_\odot$, $M_{\rm NS/BH}=1.6$\, M$_\odot$, and $\cos i
= 0$ for SDSS 1257+5428 results in a gravitational wave (GW) strain
amplitude of $h \approx 2.3 \times 10^{-21}$, which exceeds the nominal
sensitivity of LISA of $\sim 10^{-21}$ at $f_{\rm GW} \sim 0.1$~mHz (Roelofs
et al.\ 2007).  The estimate of equation (\ref{rate}) suggests that WD-NS/BH binaries
will also contribute significantly to the GW background in the LISA band
(Badenes et al.\ 2009; Kim et al.\ 2004), which may affect the detectability
of the SDSS 1257+5428 signal. 
Importantly, individual mergers in the local universe may also be important 
GW sources for LIGO if the merger results in BH formation or if 
the massive NS merger remnant is initially rapidly rotating
(Garcia-Berro et al.~2007; Paschalidis et al.~2009). 

\noindent {\it 6.~The Lowest Mass Black Holes:}
Note that the total mass of SDSS 1257+5428 probably exceeds
the maximum mass for a NS (e.g., Lattimer \& Prakash 2007).  Thus,
if the WD was entirely accreted, this event would be accompanied 
by BH formation with $\sim2.5$\,M$_\odot$  (see Brown et al.~2001). 
The rate of WD-NS/BH mergers  estimated in equation (\ref{rate})
is large enough that the Galaxy should be littered with $\sim2-3$\,M$_\odot$
BHs; there should be $\sim10^6-10^7$ low-mass free-floating BHs throughout
the Galaxy.  

\noindent {\it 7.~The Origin of Ultra-High Energy Cosmic Rays:}
If WD-NS/BH mergers produce a BH and accretion disk 
(see Popham et al.~1998; Fryer et al.~1999b), the magnetic luminosity via the
Blandford-Znajek mechanism will
likely be large enough for production of UHECRs (e.g., Waxman \& Loeb 2009; see also
Waxman 1995).
Because the energy reservoir in WD-NS mergers 
is comparable to the energy budget of GRBs,
but the overall rate of WD-NS/BH mergers within the local GZK
volume ($\sim100$\,Mpc) is  $\gtrsim100$ times  larger (eq.~\ref{rate};
\S\ref{section:discussion}), WD-NS/BH mergers may 
dominate the UHECR budget.  The merger may also produce a 
rapidly rotating (ms spin period) NS with potentially short-lived magnetar-strength
($\sim10^{15}$\,G) magnetic fields.  In this case, the mechanism
advocated by Arons (2003) for UHECR production may obtain.
The rate needed by Arons to account for the observed UHECR
budget is close to the rate derived for WD-NS/BH mergers 
estimated here (Table \ref{table}).  
Additionally, note that the formation of
ms magnetar-like conditions via WD-NS/BH mergers alleviates the 
problem of getting UHECRs out of the overlying dense massive 
star progenitor discussed by Arons.

\noindent {\it 8.~Binaries in Type Ib/c SNe:}
If the scenario outlined by Davies et al.~(2002), which predicts
a formation rate of  SDSS 1257+5428-like binaries within a factor of
$\sim1-10$ of the estimate in equation (\ref{rate}), is correct,
then the star that produces the NS (originally the secondary)
explodes after becoming a He star and transferring a significant fraction of
its envelope to the WD, and potentially expelling it from the system. The
fact that the overall rate of Type-Ib/c SNe is $\sim10-20$\% of
core-collapse SNe (Prieto et al.~2008) 
implies that if the Davies et al.~mechanism
is correct, and if $\Gamma_{\rm MW}$ is correct, then many
Type Ib/c SNe  ($\sim20-50$\%) explode with a very 
close WD companion.  

\noindent {\it 9.~Tight Companion Interaction:}
Regardless of the binary formation channel, the 
SN explosion that produces the NS may interact with 
the close secondary, be it a WD, main sequence star, or otherwise. 
In this case, there 
may be a signal of interaction in the very early-time lightcurve of 
many stripped envelope Type-Ib/c SNe as a result of the break-out 
flash and shockwave interacting with the nearby companion 
(see, e.g., Marietta \& Burrows 2000; Kasen 2009).

\noindent {\it 10.~Star \& Binary Formation:}
The Galactic birth rate of massive stars is approximately equal to the
core-collapse SN rate.  Thus, equation (\ref{rate}) implies that $\sim2 - 4$\% of all
massive stars are born with a close binary companion capable of producing a
SDSS 1257+5428-like binary.  If true, this has important implications for the
physics of massive star formation and the demographics of young massive star
binaries (e.g., Krumholz et al.\ 2009), as well as for pulsar binaries.  The
low eccentricity of the orbit of SDSS 1257+5428 suggests interaction with the
companion subsequent to the NS birth.  Additionally, if WD production precedes the 
NS, equation (\ref{rate})
implies a significant number of young pulsars with very tight WD companions.
However, only one such system has been detected among the $> 1000$
non-millisecond pulsars (MSPs) --- the highly-eccentric PSR J1141-6545 (see Edwards \&
Bailes 2001; Davies et al.\ 2002; Kim et al.\ 2004) --- disfavoring this as an
analog.  If the NS preceded the WD and was recycled into a MSP,
then MSPs should be embedded in the centers of a fraction of young planetary
nebulae, potentially visible either in radio and/or X-rays and gamma-rays.

Indeed, the estimate of equation (\ref{rate}) is large enough that the simplest
explanation for the discovery of SDSS 1257+5428 is either that SWARMS was very
lucky, or that instead of a WD-NS/BH binary, this system is a tight WD-WD binary
(Badenes et al.\ 2009).  Although the high nominal space velocity of the system
argues against this possibility, if SDSS 1257+5428 is in fact a WD-WD binary, then 
the similarity between the derived rate and
the observed Type Ia SN rate is not a coincidence (see Table~\ref{table}),
particularly in light of the results of Mullally et al.\ (2009) (see also Kilic
et al.\ 2009). The relatively high value for the rate in equation (\ref{rate})
is yet more perplexing if the companion to the observed WD is in fact a BH, and
not a NS (see Brown et al.\ 2001).  Nevertheless, the relative lack of
pulsars with tight binary companions may argue for a BH companion, and it
should be kept in mind that the SWARMS survey is perhaps the first where 
such tight WD-BH binaries could have been detected.  
Clearly, a more complete census and analysis --- as will be
provided by the SWARMS survey (Mullally et al.\ 2009) --- and follow-up
observations of SDSS 1257+5428, including a search for (potentially pulsed)
radio and X-ray emission as well as a parallax measurement, are needed.  These
will be vital, since, if SDSS 1257+5428 is radio-quiet, this would imply a class of
binaries distinct from any presently known.

\vspace{-.6cm}

\acknowledgments 

We thank C.~Badenes and  B.~Metzger for a critical reading of the text and
for comments that improved the manuscript. We thank 
J.~Beacom, C.~Kochanek, B.~Lacki, P.~Martini, O.~Pejcha, J.~Prieto, E.~Quataert, and 
D.~Zhang for additional discussions.  
T.A.T.~is supported in part by an Alfred P.~Sloan Foundation Fellowship.
M.D.K.~is supported in part by an OSU Presidential Fellowship and 
by NSF CAREER grant PHY-0547102.
T.A.T.~and K.Z.S.~are supported in part by NSF grant AST-0908816,

%------------------------------------------------------------------------------


\begin{thebibliography}{}

\bibitem[Badenes et al.(2009)]{2009arXiv0910.2709B} 
Badenes, C., et al.\ 2009, arXiv:0910.2709 

\bibitem[Bessel(1838)]{1838AN.....16...65B} 
Bessel, F.~W.\ 1838, Astronomische Nachrichten, 16, 65 

\bibitem[Brown et al.(2001)]{2001ApJ...547..345B} Brown, G.~E., et al.\ 2001, \apj, 547, 345 

\bibitem[Chen \& Beloborodov(2007)]{2007ApJ...657..383C} 
Chen, W.-X., \& Beloborodov, A.~M.\ 2007, \apj, 657, 383 

\bibitem[Davies et al.(2002)]{2002MNRAS.335..369D} 
Davies, M.~B., Ritter, H., \& King, A.\ 2002, \mnras, 335, 369 

\bibitem[Dilday et al.(2008)]{2008ApJ...682..262D} 
Dilday, B., et al.\ 2008, \apj, 682, 262

\bibitem[Edwards \& Bailes(2001)]{2001ApJ...547L..37E} 
Edwards, R.~T., \& Bailes, M.\ 2001, \apjl, 547, L37 

\bibitem[Filippenko et al.(1992)]{1992AJ....104.1543F} Filippenko, A.~V., 
et al.\ 1992, \aj, 104, 1543 

\bibitem[Foley et al.(2009)]{2009AJ....138..376F} Foley, R.~J., et al.\ 
2009, \aj, 138, 376 

\bibitem[Freiburghaus et al.(1999)]{1999ApJ...525L.121F} Freiburghaus, C., 
Rosswog, S., \& Thielemann, F.-K.\ 1999, \apjl, 525, L121 

\bibitem[Fryer et al.(1999)]{1999ApJ...526..152F} 
Fryer, C.~L., Woosley, S.~E., \& Hartmann, D.~H.\ 1999a, \apj, 526, 152 

\bibitem[Fryer et al.(1999)]{1999ApJ...520..650F} 
Fryer, C.~L., et al.\ 1999b, \apj, 520, 650 

\bibitem[Gal-Yam et al.(2006)]{2006Natur.444.1053G} 
Gal-Yam, A., et al.\ 2006, \nat, 444, 1053 

\bibitem[Garcia-Berro et al.(2006)]{garcia}
Garcia-Berro et al.~2007, Journal of Physics Conference Series, 66, 012040

\bibitem[Gehrels(1986)]{1986ApJ...303..336G} Gehrels, N.\ 1986, \apj, 303,  336 

\bibitem[Gehrels et al.(2006)]{2006Natur.444.1044G} 
Gehrels, N., et al.\ 2006, \nat, 444, 1044 

\bibitem[Guetta et al.(2005)]{2005ApJ...619..412G} 
Guetta, D., Piran, T., \& Waxman, E.\ 2005, \apj, 619, 412 

\bibitem[Horiuchi et al.(2009)]{2009PhRvD..79h3013H} 
Horiuchi, S., Beacom, J.~F., \& Dwek, E.\ 2009, \prd, 79, 083013 

\bibitem[Kalogera et al.(2001)]{2001ApJ...556..340K} 
Kalogera, V., et al.\ 2001, \apj, 556, 340 

\bibitem[Kalogera et al.(2004)]{2004ApJ...601L.179K} 
Kalogera, V., et al.\ 2004a, \apjl, 601, L179 

\bibitem[Kalogera et al.(2004)]{2004ApJ...614L.137K} 
Kalogera, V., et al.\ 2004b, \apjl, 614, L137 

\bibitem[Kalogera et al.(2005)]{2005ASPC..328..261K} 
Kalogera, V., et al.\ 2005, Binary Radio Pulsars, 328, 261 

\bibitem[Kasen(2009)]{2009arXiv0909.0275K} 
Kasen, D.\ 2009, arXiv:0909.0275 

\bibitem[Kawabata et al.(2009)]{2009arXiv0906.2811K} 
Kawabata, K.~S., et al.\ 2009, arXiv:0906.2811 

\bibitem[Kilic et al.(2009)]{2009arXiv0911.1781K} 
Kilic, M., et al.\ 2009, arXiv:0911.1781 

\bibitem[King et al.(2007)]{2007MNRAS.374L..34K} 
King, A., Olsson, E., \& Davies, M.~B.\ 2007, \mnras, 374, L34 

\bibitem[Krumholz et al.(2009)]{2009Sci...323..754K} 
Krumholz, M.~R., et al.\ 2009, Science, 323, 754 

\bibitem[Lattimer \& Prakash(2007)]{2007PhR...442..109L} 
Lattimer, J.~M., \& Prakash, M.\ 2007, \physrep, 442, 109 

\bibitem[Li et al.(2000)]{2000AIPC..522..103L} Li, W.~D., et al.\ 2000, 
American Institute of Physics Conference Series, 522, 103 

\bibitem[Marietta et al.(2000)]{2000ApJS..128..615M} Marietta, E., Burrows, 
A., \& Fryxell, B.\ 2000, \apjs, 128, 615 

\bibitem[Metzger et al.(2007)]{2007ApJ...659..561M} 
Metzger, B.~D., Thompson, T.~A., \& Quataert, E.\ 2007, \apj, 659, 561 

\bibitem[Metzger et al.(2008)]{2008ApJ...676.1130M} 
Metzger, B.~D., Thompson, T.~A., \& Quataert, E.\ 2008, \apj, 676, 1130 

\bibitem[Metzger et al.(2009)]{2009MNRAS.396.1659M} 
Metzger, B.~D., Piro, A.~L., \& Quataert, E.\ 2009, \mnras, 396, 1659 

\bibitem[Meyer(2002)]{2002PhRvL..89w1101M} 
Meyer, B.~S.\ 2002, Physical Review Letters, 89, 231101 

\bibitem[Mullally et al.(2009)]{2009arXiv0911.1798M} 
Mullally, F., et al.\ 2009, arXiv:0911.1798 

\bibitem[Nakar(2007)]{2007PhR...442..166N} 
Nakar, E.\ 2007, \physrep, 442,  166 

\bibitem[Nelemans et al.(2001)]{nelemans}
Nelemans, G., Yungelson, L.~R., \& Portegies Zwart, S.~F.~2001, A\&A, 375, 890

\bibitem[Norris et al.(2009)]{2009arXiv0910.2456N} 
Norris, J.~P., Gehrels, N., \& Scargle, J.~D.\ 2009, arXiv:0910.2456 

\bibitem[Paschalidis et al.(2009)]{2009PhRvD..80b4006P} 
Paschalidis, V., et al.\ 2009, \prd, 80, 024006 

\bibitem[Perets et al.(2009)]{2009arXiv0906.2003P} 
Perets, H.~B., et al.\ 2009, arXiv:0906.2003 

\bibitem[Popham et al.(1999)]{1999ApJ...518..356P} 
Popham, R., Woosley, S.~E., \& Fryer, C.\ 1999, \apj, 518, 356 

\bibitem[Poznanski et al.~(2009)]{poznanski}
Poznanski, D.~et al.~2009, Science, 118, 1709

\bibitem[Phinney(1991)]{1991ApJ...380L..17P} Phinney, E.~S.\ 1991, \apjl,  380, L17 

\bibitem[Prieto et al.(2008)]{2008ApJ...673..999P} Prieto, J.~L., Stanek, 
K.~Z., \& Beacom, J.~F.\ 2008, \apj, 673, 999 

\bibitem[Roelofs et al.(2007)]{2007ApJ...666.1174R} 
Roelofs, G.~H.~A., et al.\ 2007, \apj, 666, 1174 

\bibitem[Stanek et al.(2006)]{2006AcA....56..333S} 
Stanek, K.~Z., et al.\ 2006, Acta Astronomica, 56, 333 

\bibitem[Valenti et al.(2009)]{2009Natur.459..674V} 
Valenti, S., et al.\ 2009, \nat, 459, 674 

\bibitem[Waxman \& Loeb(2009)]{2009JCAP...08..026W} 
Waxman, E., \& Loeb, A.\ 2009, JCAP, 8, 26 

\bibitem[Waxman(1995)]{1995PhRvL..75..386W} Waxman, E.\ 1995, Physical 
Review Letters, 75, 386 

\end{thebibliography}
\end{document}